\documentclass[letterpaper, 10 pt, conference]{ieeeconf} 

\IEEEoverridecommandlockouts                              

\usepackage{graphics} 
\usepackage{epsfig} 
\usepackage{mathptmx} 
\usepackage{times} 
\usepackage{amsmath} 
\usepackage{amssymb}  
\usepackage{algorithm} 
\usepackage{algpseudocode}
\usepackage{cite}
\usepackage{url}
\usepackage{hyperref}
\hypersetup{breaklinks=true}
\usepackage{svg}
\def\BibTeX{{\rm B\kern-.05em{\sc i\kern-.025em b}\kern-.08em
    T\kern-.1667em\lower.7ex\hbox{E}\kern-.125emX}}

\title{\LARGE \bf
Conflict-Free Flight Scheduling Using Strategic Demand Capacity Balancing for Urban Air Mobility Operations
}

\author{Vahid Hemmati$^{1}$, Yonas Ayalew$^{1}$, Ahmad Mohammadi$^{1}$, Reza Ahmari$^{1}$, Parham Kebria$^{1}$,\\ Abdollah Homaifar$^{1*}$, and Mehrdad Saif$^{2}$
\thanks{$^{1}$Authors are with the Department of Electrical and Computer Engineering at North Carolina A\&T State University, Greensboro, NC 27411, USA.}
\thanks{$^{2}$M. Saif is with the Department of Electrical and Computer Engineering, Windsor University, Windsor, ON N9B 3P4, Canada.}
\thanks{This research is supported by the National Aeronautics and Space Administration University Leadership Initiative (NASA-ULI 2019) through grant number 80NSSC20M0161, and partially by the University Transportation Center (UTC) USA through grant number 69A3552348327.}
\thanks{$^*$Corresponding author (homaifar@ncat.edu).}
}

\begin{document}

\maketitle
\thispagestyle{empty}
\pagestyle{empty}

\begin{abstract}
In this paper, we propose a conflict-free multi-agent flight scheduling that ensures robust separation in constrained airspace for Urban Air Mobility (UAM) operations application. First, we introduce Pairwise Conflict Avoidance (PCA) based on delayed departures, leveraging kinematic principles to maintain safe distances. Next, we expand PCA to multi-agent scenarios, formulating an optimization approach that systematically determines departure times under increasing traffic densities. Performance metrics, such as average delay, assess the effectiveness of our solution. Through numerical simulations across diverse multi-agent environments and real-world UAM use cases, our method demonstrates a significant reduction in total delay while ensuring collision-free operations. This approach provides a scalable framework for emerging urban air mobility systems.

\end{abstract}

\section{INTRODUCTION}

Urban Air Mobility (UAM) enables highly automated, cooperative, passenger, or cargo-carrying air transportation services in and around urban areas~\cite{fontaine2023urban}. The potential emergence of large-scale UAM operations poses a number of challenges for secure, safe and effective Air Traffic Management (ATM) \cite{ahmadicaic, ahmadsmc, ahmadifsa, ahmadvehicular}. Secure and safe traveling requires reliable perception from the airspace environment\cite{rezaifsDD, rezaifstrojan, p3, rezasmc, rezamynudd, chowdhury2024performance}. Operations of UAM vehicles are based on the emerging features that include separation automation to ensure flight path safety through cooperative and self-enabled separation assurance. The International Civil Aviation Organization (ICAO) Global ATM Operations Concept provides a structure through
which autonomous operations can be established with three
layers of conflict management: strategic conflict management,
separation provision, and collision avoidance~\cite{icao_document}. Strategic conflict management is generally considered as the first layer of conflict management for safe flight operations to condition the traffic to reduce the need for airborne separation provision, the second layer of conflict management. Strategic conflict management involves strategic decisions like ground delays made by air traffic controllers to balance traffic demand with airspace capacity at vertiports, merging points, and intersections~\cite{chen2024integrated}. 

Before an aircraft enters an airspace, the flight operator calculates and submits its preferred estimated trajectory traversing the airspace. Since trajectories are planned individually and independently, they can cause conflict or congestion, leading to a high system-wide cost~\cite{qian2017coordinated}. The ATM system, or air traffic controllers, needs to coordinate among the received trajectories, carrying out necessary modifications, including ground delay programs (GDP), to ensure the safety of air traffic and improve its efficiency, with the help of decision support tools. Regarding air traffic safety, en-route aircraft are required to keep separation from each other throughout their flight. Two aircrafts are defined to be in-conflict if their relative distance is less than a given minimum separation boundary both horizontally and vertically. 

In this work, we consider a pairwise strategic conflict management problem in the en-route phase of UAM operation, and aim to provide conflict-free scheduling of aircrafts using ground delay respecting the safety buffer of each UAM aircrafts. The conflict management strategy computes the ground delay for each aircraft in a multi-agent system considering optimization problem in a two-dimensional shared airspace. Average flight delay time performance metrics is used to assess the performance of the conflict management strategy with different number of multi-agent systems. The main contributions of the paper are summarized below:


\textbf{Pairwise Conflict Avoidance Framework}: We introduced a novel pairwise conflict avoidance method based on strategic demand capacity balancing for UAM operations, ensuring safe flight separation in constrained airspace. 
\textbf{Optimization of Flight Scheduling}: The paper presents an optimization approach to determine the optimal departure times for multiple UAM aircraft, minimizing delays while ensuring conflict-free operations. This strategy accommodates increasing traffic densities in real-world scenarios. 
\textbf{Scalable Multi-Agent Solution}: We expanded the PCA framework to multi-agent scenarios, demonstrating the scalability of our approach by testing various traffic densities and providing a generalized framework for UAM systems with different operational requirements.
\textbf{Simulation-Based Performance Evaluation}: The proposed method was validated through numerical simulations in different multi-agent settings and a case study based on potential UAM routes in the Greater Atlanta Metro Area. Our method significantly reduced delays while ensuring robust conflict avoidance.
\textbf{Real-World Application}: A case study was conducted in a real-world environment, showing how our method can be applied to optimize flight schedules and reduce delays in UAM operations within an urban setting.

The remainder of this paper is structured as follows: Section
II discusses ideas and approaches covered in related research.
Section III provides details of the proposed methods used in
the development of the scheduling algorithm. Section IV presents the results and discussions of the conflict management strategy in multi-agent system and use case scenario in Atlanta metro area potential routes. Section V summarizes the work and provides its conclusion and finally, section VI provides future directions of the work.

\section{Related Works}

Various studies have explored safety in autonomous robotics, aviation, and UAM systems~\cite{ayalew2023data, tereda2024efficient, p2, hemmati2024mission, ayalew2024flight, p1, tereda2023predictive, nuhu2023local}. In~\cite{lee2022demand}, a Demand Capacity Balancing (DCB) algorithm is proposed for UAM operations to strategically deconflict flights at vertiports by delaying departures when capacity is exceeded. This aggregate-level method ensures the number of flights at specific vertiport resources does not surpass their capacity.

Extending this idea,~\cite{moolchandani2024demand} introduces a DCB algorithm for enroute waypoints. It sequentially selects flights and resolves imbalances at constrained nodes—whether vertiports or enroute conflict points. The effectiveness is assessed by the number and magnitude of delays imposed.

Conflict resolution has also been addressed in several works. In~\cite{qian2017coordinated}, UAVs are assigned predefined tracks and speeds to eliminate potential conflicts through scheduled takeoffs. A hybrid model in~\cite{shihab2019schedule} integrates on-demand and scheduled services for UAM, leveraging mixed-integer quadratic programming for efficient conflict management. For terminal areas,~\cite{bertram2020efficient} introduces a method combining loops, gates, and a terminal sorting algorithm using a Markov decision process to reduce delays in dense UAM environments.

A more integrated approach is presented in~\cite{pang2022adaptive}, where conflict resolution is formulated as an optimization problem involving scheduling, speed control, and rerouting. In~\cite{pang2024machine}, a machine-learning-based method is developed for aircraft landing scheduling to improve safety and deconfliction using data-driven models.

While the literature has advanced strategic DCB and optimization techniques, real-time conflict resolution under high-density UAM conditions remains insufficiently addressed. To bridge this gap, our work proposes a pairwise enroute conflict resolution strategy using ground delays, optimizing schedules to minimize total system delay. We evaluate the method’s scalability using multi-agent scenarios and apply it to a case study on urban UAM routes.

\section{Methodology}
Our proposed methodology consists of three stages: (A) Introducing Pairwise Conflict Avoidance (PCA) by Delay, (B) Expanding PCA to Multi-Agent Scenarios, and (C) Establishing Performance Metrics.

\subsection{Introducing Pairwise Conflict Avoidance (PCA) by Delay}

Consider a scenario that involves two agents, Agent A and Agent B, operating within a two-dimensional airspace. Initially, both agents are stationed at their respective positions $\vec{P_{A}}^{\circ}$ and $\vec{P_{B}}^{\circ}$. The agents are scheduled to commence their missions sequentially, with agent A starting first, followed by agent B after a delay of $t_d$.

Agent A is already in motion, following a predefined path at a constant velocity $\vec{V}_{A}$. On the other hand, agent B, also moving at a constant velocity $\vec{V}_{B}$, must avoid encroaching upon agent A's path. To ensure safe separation between the two agents, a circular buffer zone with a radius of $h$ is defined around agent A. Any violation of this buffer zone by agent B could lead to a potential collision or conflict situation.

To prevent such conflicts, it is crucial to calculate a safe delay time $t_d$ that allows agent B to depart from its initial position without breaching the buffer zone around agent A. This calculation involves considering the velocities and initial positions of both agents, as well as the size of the buffer zone.

In this section, we aim to derive an expression for the safe delay time $t_d$ that guarantees the avoidance of conflicts between agents A and B. This analysis will be  the fundamental principles of kinematics and collision avoidance for pair agents. Equation \ref{eqn:eq001} represents the primary problem statement of the current section.
\begin{figure}[h!]
\centering
\includegraphics[width=0.45\columnwidth]{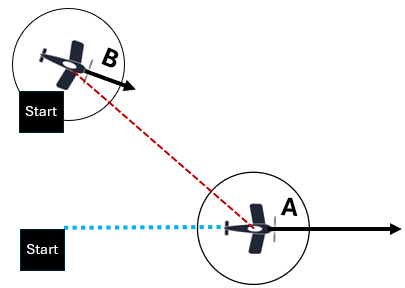}
\caption{Agent A starts $t_d$ units before Agent B. At Agent B's start, A has followed the blue path. The red dashed line shows $R(t)$, which must stay above the minimum separation $h$.}

\label{fig:pic01}
\end{figure}
\begin{equation}
     Find:~t_d \ni  \left | \vec{P}_A(t+t_d) - \vec{P}_{B}(t) \right |\geq h
    \label{eqn:eq001}
\end{equation}

Equation \ref{eqn:eq002} describes the relative position of agent A with respect to agent B as follows:

\begin{equation}
    \vec{R}(t)= (\vec{V}_A-\vec{V}_B)~t + (\vec{P}_A-\vec{P_B^{\circ}})
    \label{eqn:eq002}
\end{equation}

Given that $\vec{P}_A= \vec{P_A^{\circ}} + \vec{V}_A~t_d $ we simplify the notation by introducing the relative velocity U and relative position P. Therefore, Equation \ref{eqn:eq002} can be expressed as follows:

\begin{equation}
    \vec{R}(t)= \vec{U}t + \vec{P}
    \label{eqn:eq003}
\end{equation}


Where, $\vec{U}= (\vec{V}_A-\vec{V}_B)$ and $\vec{P}=(\vec{P}_A-\vec{P_B^{\circ}})$ represent the relative velocity and relative position at the departure time of agent B and are assumed to be constant. The squared distance $|\vec{R}|^{2}$ between the agents, which must always remain greater than $h^2$ , is calculated using Equation \ref{eqn:eq004}.

\begin{equation}
\begin{split}
    \left |\vec{R}(t)\right|^{2}&= (\vec{U}t+\vec{P})^{T}.(\vec{U}t + \vec{P})\\
    &=|\vec{U}|^{2}t^{2}+ 2(\vec{U}.\vec{P})t+ |\vec{P}|^{2}
    \label{eqn:eq004}
\end{split}    
\end{equation}


In the context of two agents, A and B, moving in a two-dimensional space, the minimum separation distance between them must be maintained to ensure safety. The separation distance, denoted as 
$h$, represents the radius of a circular buffer zone around each agent. No other agent should enter this zone to prevent potential collisions.


To find the time \( t_{\text{min}} \) when the minimum separation distance occurs, we examine how the squared distance between the agents, \( |\vec{R}|^2 \), changes over time. The derivative of \( |\vec{R}|^2 \) with respect to time, \( \frac{\partial |\vec{R}(t)|^2}{\partial t} \big|_{t = t_{\text{min}}} = 0 \), equals zero when \( |\vec{R}|^2 \) reaches its minimum.

Once we found $t_{min}$, we then solve Equation \ref{eqn:eq004} to ensure that the distance between the agents is exactly $h$ at $t_{min}$. This calculation helps us determine the exact positions of agents A and B at the point of minimum separation, allowing us to plan their trajectories to avoid any potential conflicts.

\begin{equation}
\min(R^2)= R^2|_{t=t_{min}}= h^2
\label{eqn:eq005}
\end{equation}

Finally,  Eq. \ref{eqn:eq006} should be solved for finding the delay time, $t_d$: 

\begin{equation}
\frac{(\Vec{U}.\Vec{P})^2}{|\vec{U}|^2} +|\Vec{P}|^2 = h^2
\label{eqn:eq006}
\end{equation}


Substituting the values of $\vec{P}=(\vec{P}_A-\vec{P_B^{\circ}})$ and $\vec{U}=(\vec{V}_A-\vec{V}_B)$  into Equation \ref{eqn:eq006}, we obtain Equation \ref{eqn:eq007}.

\begin{equation}
\centering
\begin{split}
h^2=\frac{1}{|\vec{V}_A-\vec{V}_B|^2} |(\vec{V}_A-\vec{V}_B).(\vec{P_A^{\circ}}+t_d.\vec{V}_A-\vec{P_B^{\circ}})|^2\\
+|\vec{P_A^{\circ}}+t_d.\vec{V}_A-\vec{P_B^{\circ}}|^2
\label{eqn:eq007}
\end{split}
\end{equation}

\begin{figure}
\centering
\includegraphics[width=0.5\columnwidth]{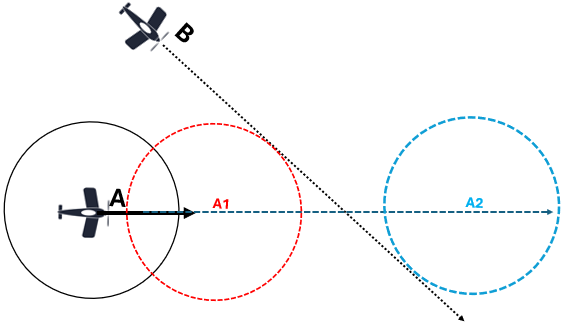}
\caption{The two roots of Eq.~\ref{eqn:eq007} represent two cases: a longer delay lets Agent A reach A2 (blue) before being overtaken; a shorter delay causes Agent B to catch A at A1 (red).}

\label{fig:pic02}
\end{figure}

Equation \ref{eqn:eq007} is a quadratic equation with respect to the delay time, $t_d$. Generally, such an equation has two roots, and the interval between the roots represents the time span during which the separation limit is violated. Figure \ref{fig:pic02} illustrates the scenario when both roots are positive.
The physical interpretation of negative roots signifies a reversal in the departure order. In other words, agent A needs to wait for agent B to depart first, contrary to the assumption made when formulating the equations.

\subsection{Expanding PCA to Multi-Agent Scenarios}

\begin{figure}
\centering
\includegraphics[width=0.5\columnwidth]{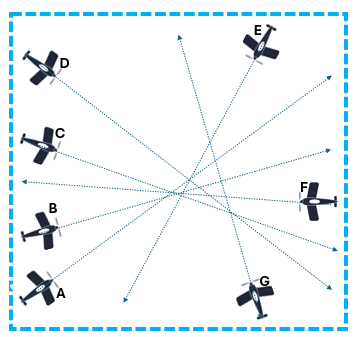}
\caption{In the general case with $N = 7$ agents, there are $N!$ possible departure orders. Managing conflict-free timings and optimizing the sequence to minimize delays are key challenges.}
\label{fig:pic03}
\end{figure}

In real scenarios, a list of requested flights is represented as $\vec{F}=[A,B,C ...,G]$ as depicted in Figure \ref{fig:pic03}. The departure ports for these flights are known. However, determining the exact departure times, $\vec{T}=[t^{d}_A, t^{d}_B, ..., t^{d}_G ]$ can be an optimization problem aimed at resource efficiency.

This section elaborates on how the proposed PCA method can be extended to handle multi-agent scenarios. The goal is to minimize the total delay, $\min(\sum_{i}^{N}t_i^d)$ while ensuring that no conflicts occur between flights. A conflict is defined as $ |\vec{P}_{F_i}(t+t_i^d)-\vec{P}_{F_j}(t+t_j^d)|< h$, where $h$ is the minimum separation distance. Equation \ref{eqn:eq008} represents the main objective of this section, which is to optimize the departure times to minimize delays while preventing conflicts.

\begin{equation}
   \arg \min_{\vec{T}} (\sum_{i}^{N}t_i^d \ni |\vec{P}_{F_i}(t+t_i^d)-\vec{P}_{F_j}(t+t_j^d)|\geq h ) 
   \label{eqn:eq008}
\end{equation}

In the optimization process for scheduling multiple agents' departures, ensuring that each agent can depart without conflicting with others that are already in the air is crucial. This process involves iteratively determining the earliest feasible departure time for each agent. Algorithm \ref{algorithm:1} describes the procedure as follows:
\textbf{Conflict Avoidance:} At each agent's turn, the algorithm considers all possible departure times span within the available total time span (total time span is from the first departure to the last arrival in the flight list) and eliminates the time spans that could cause conflicts with agents that are already in the air. This ensures that each agent's departure time is selected such that there is no conflicts.
\textbf{Reserving the Earliest Feasible Time:} After eliminating conflicting departure times, the algorithm reserves the earliest feasible departure time within the collision-free time spans for each agent. This reservation aims to minimize the overall delay in the system by allowing agents to depart as soon as possible without any conflicts.
\textbf{Input and Output:} The algorithm takes the list of flights in order, $\vec{F}=[A,B,C ...,G]$ as its input. It then generates the list of departure times, $\vec{T}=[t^{d}_A, t^{d}_B, ..., t^{d}_G ]$, as its output. Each departure time, $t^{d}_i$, in the output list is relative to the departure time of the first agent in the flight list, $F$.
\textbf{Horizon Time:} The horizon time, $H$, represents the time interval starting from the departure of the first agent ($t=0$) to the arrival time of the last agent ($t=t_{final}$). While this interval ideally covers the entire duration of all flights, an overestimated upper bound, $H$, is used in practice to ensure that all possible conflicts are considered.

By applying this optimization process, algorithm \ref{algorithm:1} seeks to schedule the departure times of multiple agents in a manner that minimizes delays and prevents conflicts, ensuring efficient and safe operations in the airspace.

\begin{algorithm}
	\caption{Algorithm for expanding PCA to Multi-Agents} 
	\begin{algorithmic}[1]
\State $F \gets  List~of~flights~in~order $
\State $T \gets  List~of~zeros~with~length~of~ F$
\State $H \gets  Horizon~time:~[~0,~t_{final}]$
		\For {$agent_j~ in ~  F$}
                \For {$agent_i~ in ~ F ~ befor~agent_j$}
                \State $ Confilict \gets PCA(agent_i,~agent_j) $
                \State $H \gets  (H~-~Confilict)$
                \EndFor
                \State $T[index~(agent_j)] =  Min(H)$
			\EndFor
\State $Return~ Output:~ T$ 
	\end{algorithmic} 
    \label{algorithm:1}
\end{algorithm}


\subsection{Establishing Performance Metrics}
Given the objective outlined in Equation \ref{eqn:eq008}, it is essential to introduce a metric that assesses the proposed method with respect to time. In practical scenarios, delays incur additional costs for airline service providers. The total delay time can be quantified as the sum of all elements in the departure time list, $\vec{T}$. However, to compare delay times across airspaces with varying air traffic densities, the average delay time serves as a more suitable metric, as shown in Equation \ref{eqn:eq009}.

\begin{equation}
   \bar{t^d} = \frac{1}{N}\sum_{i=1}^{N}t_{i}^d 
   \label{eqn:eq009}
\end{equation}
Where $N$ is the number of agents in the flight list, $\vec{F}$.    


To assess the effectiveness of the proposed method, we developed a scaled-down simulation environment using Python. This environment models an airspace with dimensions of 20×20 meters. Within this space, vertiports are randomly positioned along the airspace's circumference. To maintain safe separation between agents, a minimum separation limit of $h=1.5$ meters is defined.

In this simulation, the number of agents varies from 2 to 7, allowing us to observe how the proposed method performs under different air traffic congestion. The missions' velocities are randomly assigned from a range of 0.66 to 1.89 meters per second. This velocity range was chosen for two main reasons.

\textbf{Firstly}, it ensures that the travel times of the missions fall within a reasonable interval of 15 to 30 seconds. This range is important for visualization purposes, as it allows us to observe the movements of the agents in a meaningful time frame (not too long).

\textbf{Secondly}, this time interval is also considered reasonable for defining a collision window. A collision window is the time interval during which a collision between two agents could occur if their paths intersect (not too short). 

By keeping the travel times within this range, we can manage the computational costs associated with determining potential collisions between agents.

Overall, this simulation environment allows us to assess the proposed method's performance in a controlled setting that reflects real-world conditions in terms of airspace dimensions, traffic density, and mission velocities.



The performance of PCA method has demonstrated that the roots of Equation \ref{eqn:eq007} yield optimized trajectories, allowing agents to pass tangentially to the safety buffer, as illustrated in Figure \ref{fig:pic02}. Additionally, it was observed that negative roots lead to a reversal in the order of departures, as anticipated.

Subsequently, multi-agent simulations were conducted for the airspace, varying the number of agents from 4 to 7. Details of these experiments are discussed in the following section.

\section{Simulation Results}

This section demonstrates the application of the suggested method, which can be utilized to generate baseline data for future developments. The following subsections discuss the results of two experiments: (A) The Effect of Air Traffic Density (ATD) on Departure Delays, and (B) A Case Study in the Greater Atlanta Metro Area, USA.

\subsection{Effect of ATD on Departure Delays}


The simulation environment covers an area of $400 [m^2]$ and supports varying numbers of agents, with $N$ ranging from 4 to 7. Establishing the mission topology requires $2N$ vertiports, randomly positioned along the circumference of the square airspace. The missions are assigned randomly to ensure that the nominal trajectories always intersect. The total number of flight orders is given by $N!$. For instance, a 4-agent simulation involves $4! = 24$ flight order combinations, while a 7-agent simulation has $7! = 5,040$ combinations.
To maintain a consistent number of flights (approximately 120,000 missions) for each ATD scenario, different numbers of topologies, denoted as $TN$, have been tested for $N = {4, 5, 6, 7}$, with $TN$ values of {5000, 1000, 166, 24}, respectively.

\begin{figure}
\centering
\includegraphics[width=0.8\columnwidth]{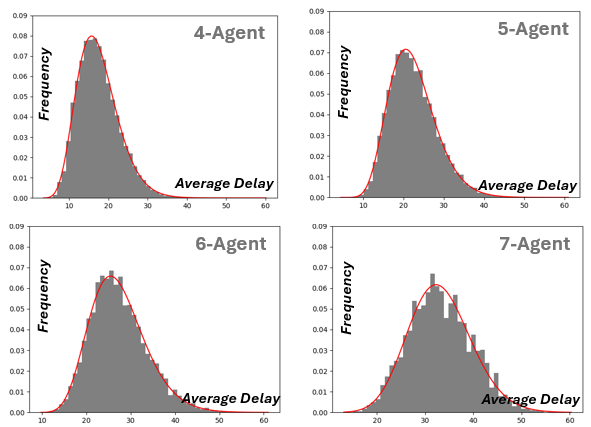}
\caption{Gray histograms show average delays for different ATDs. The red curve is the best-fit Gamma distribution with the lowest SSR.}
\label{fig:pic04}
\end{figure}


The simulation produced departure times, $\vec{T}$, and Equation \ref{eqn:eq009} was utilized to compute the average delay. As illustrated in Figure \ref{fig:pic04}, histograms displaying the average delay were plotted. The distribution function with the lowest Sum of Squared Residuals (SSR) was identified as the best fit.

\begin{table}[ht]
\caption{SSR for each Distribution Function} 
\centering 
\begin{tabular}{c c c c c} 
\hline\hline 
Case: & Normal & Log-norm & Beta &  Gamma \\ [0.5ex] 
\hline 
SSR:  & 0.04507478 & 0.031768467 & 0.03234198 & 0.03166473\\ [1ex] 
\hline 
\end{tabular}
\label{table01} 
\end{table}

\begin{figure}
\centering
\includegraphics[width=0.6\columnwidth]{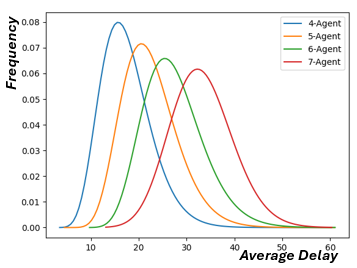}
\caption{Gamma distributions show that both average delay and standard deviation increase with ATD, highlighting greater sensitivity to departure timing at higher densities.}
\label{fig:pic05}
\end{figure}

The results presented in Table \ref{table01} indicate that Gamma and log-normal distributions offer the best fit with the lowest Sum of Squared Residuals (SSR) for the data. The Gamma distributions have been plotted in Figure \ref{fig:pic05}. The graph illustrates that higher ATDs lead to increased average delay. However, the standard deviation (StD) also increases with higher ATDs, suggesting that managing departure times has a more pronounced impact at higher ATDs. Effective management strategies can result in significant time savings and reduced delays at higher ATDs. This trend is further supported by the average and StD values reported in Table \ref{table02}. Notably, finding an optimized combination to minimize delay is crucial at higher ATDs and is one of the key outcomes of the simulation.

 \begin{table}[b]
\caption{Delay in Different ATDs} 
\centering 
\begin{tabular}{c c c } 
\hline\hline 
~ & Average Delay & Standard Deviation \\ 
\hline 
4-Agent  & 17.40795 & 5.25161 \\
5-Agent  & 22.36984 & 5.82152 \\
6-Agent  & 27.29532 & 6.29428 \\
7-Agent  & 33.14910 & 6.51092 \\ 
\hline 
\end{tabular}
\label{table02} 
\end{table}

\subsection{Case Study, Greater Atlanta Metro Area, USA }

\begin{figure}
\centering
\includegraphics[width=0.55\columnwidth]{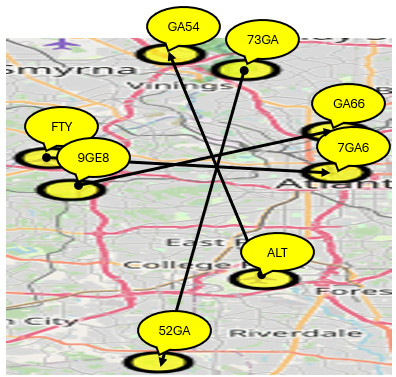}
\caption{
The topology of the selected vertiports from Atlanta for the case study is illustrated. The missions are indicated by black arrows.}
\label{fig:pic011}
\end{figure}

Eight vertiports in the Greater Atlanta Metro Area were selected for the simulation, as depicted in Figure \ref{fig:pic011}. These vertiports served as the starting and ending points for the missions. Among these vertiports, four missions were assigned, each with a specific Global Coordinate (GC) destination, as listed in Table \ref{table03}.

The cruise speed of each agent was randomly selected from a range of $55$ to $65$ [mph]. This range was chosen to reflect realistic cruising speeds for aerial vehicles in urban environments.


\begin{table}[ht]
\caption{Missions assigned to agents } 
\centering 
\begin{tabular}{c c c c c} 
\hline\hline 
Mission&From: & GC Departure & To: & GC Arrival  \\ 
\hline 
01&9GE8 & (33.741,-84.513) & GA66& (33.810,-84.395)\\
02&FTY  & (33.779,-84.521) & 7GA6& (33.762,-84.396)\\
03&ATL  & (33.637,-84.428) & GA54& (33.901,-84.468)\\
04&73GA & (33.883,-84.436)& 52GA2& (33.538,-84.474)\\ 
\hline 
\end{tabular}
\label{table03} 
\end{table}


After establishing the airspace topology and assigning missions, the algorithm automatically generated all possible mission orders, totaling 24 different combinations. Subsequently, the optimized schedule with the minimum average delay was selected.

The results revealed that the minimum total delay is approximately 20.6769 minutes, with an average delay of 5.1692 minutes. This represents a $55\%$ increase in efficiency compared to the maximum delay order, which results in a total delay of 46.5231 minutes. The efficiency is expected to further increase at higher ATDs, aligning with the trend observed in Figure \ref{fig:pic05}.

Table \ref{table04} shows the optimum flight order with their corresponding departure times.

\begin{table}[ht]
\caption{Optimized Flight Schedule } 
\centering 
\begin{tabular}{c c c c } 
\hline\hline 
Flight & Departure$[minute]$ & Route & Speed~$[mil/h]$   \\ [0.5ex] 
\hline 
03& 0.0 & ATL-GA54  & 62.4 \\
02& 5.2 & FTY-7GA6  & 55.5\\
01& 5.2 & 9GE8-GA66 & 60.7\\
04& 10.3 & 73GA-52GA2  & 62.4\\[1ex] 
\hline 
\end{tabular}
\label{table04} 
\end{table}


Interestingly, another flight order,$\vec{F}=[3,2,4,1]$, was found to yield the exact optimized average delay.

\section{Conclusions}
In this paper, we proposed a conflict-free multi-agent flight scheduling strategy for UAM operations. Our approach utilizes PCA based on strategic demand capacity balancing to minimize delays and prevent conflicts between UAM aircraft in constrained airspace. Through optimization of departure times and performance evaluation, we demonstrated that our method significantly reduces total delays while ensuring the safety of air traffic. The scalability of the approach was confirmed through simulations involving varying air traffic densities and a case study in the Greater Atlanta Metro Area. The results highlight the efficiency of the proposed method, particularly in high-density environments, where effective scheduling strategies can lead to significant improvements in operational efficiency.

\section{Future Work}
In future developments, the proposed algorithm will be enhanced to generate data that can be used as ground truth for data-driven model development. By providing reliable, conflict-free flight schedules and performance metrics, the algorithm will serve as a foundation for training machine learning models aimed at improving conflict resolution and flight scheduling strategies. This will help in advancing autonomous air traffic management systems for UAM applications, allowing the integration of real-time data and further optimization of scheduling decisions.


\bibliographystyle{IEEEtran}
\bibliography{ref}

\end{document}